\documentclass[11pt]{article}
\usepackage[margin=2.5cm]{geometry}
\usepackage{amsmath, amssymb, amsthm}
\usepackage{mathtools}
\usepackage{authblk}
\usepackage{graphicx}
\usepackage{float}
\usepackage{booktabs}
\usepackage{csquotes}
\usepackage{xcolor}
\usepackage[english]{babel}
\usepackage{microtype}
\usepackage{enumitem}

\usepackage{algorithm}
\usepackage{algpseudocode}

\usepackage[backend=biber,style=apa,citestyle=authoryear]{biblatex}
\usepackage[colorlinks=true,linkcolor=blue,citecolor=blue,urlcolor=blue]{hyperref}
\theoremstyle{plain}
\newtheorem{theorem}{Theorem}[section]
\newtheorem{lemma}[theorem]{Lemma}
\newtheorem{corollary}[theorem]{Corollary}
\newtheorem{prop}[theorem]{Proposition}
\theoremstyle{definition}

\theoremstyle{remark}
\newtheorem{remark}[theorem]{Remark}

\begin{document}

\title{Exact conditional goodness-of-fit tests for the mixed membership stochastic block model}

\author{Sourav Majumdar \\\texttt{souravm@iitk.ac.in}}
\affil{Department of Management Sciences, Indian Institute of Technology Kanpur, India}
\date{}
\maketitle

\begin{abstract}
We propose exact conditional goodness-of-fit tests for directed mixed membership stochastic block models.  Given dyad-level sender and receiver roles, the block-pair edge totals are sufficient for the block probability matrix; conditioning on these totals gives a nuisance-free uniform law on a finite fiber.  This yields finite-sample randomization tests for residual sender and receiver heterogeneity, reciprocity, and directed transitive closure.  The procedure uses an independent fiber sampler, Monte Carlo rank \(p\)-values, and can be applied after drawing latent block-pair assignments from the posterior distribution.  Simulations and the Sampson monastery network show that the tests are calibrated under the null and diagnostically useful for directed model misspecification.
\end{abstract}


\section{Introduction}
\label{sec:intro}

Network data often contain several kinds of heterogeneity at once.  Actors may differ in their baseline propensity to send or receive ties, groups may have different levels of within- and between-group connectivity, and directed networks may display reciprocity or transitive closure beyond what is explained by group structure alone.  Stochastic block models provide one of the central statistical descriptions of such heterogeneity.  The classical model of \textcite{holland1983stochastic} represents actors through latent blocks, while later variants allow for features such as degree heterogeneity \parencite{karrer2011stochastic}.  The mixed membership stochastic block model (MMSBM) of \textcite{airoldi2008mixed} adds a further layer of flexibility by allowing each actor to participate in several latent groups and by drawing sender and receiver roles at the dyad level.  This idea is closely related to latent Dirichlet allocation \parencite{blei2003latent} and to the broader mixed-membership modelling framework surveyed in \textcite{airoldi2014handbook}.  It is particularly natural for social, organizational, and relational data in which a node need not have a single stable community identity.

A large literature studies estimation, community recovery, and detection in block models; see, for example, the survey of \textcite{abbe2017community}.  Related hypothesis-testing questions include testing for the presence of community structure \parencite{bickel2016hypothesis}, testing sparse or weighted stochastic block models \parencite{yuan2022hypothesis}, and detecting small clusters in stochastic block models \parencite{ye2025detecting}.  These works address important global questions about block structure.  The question in the present paper is different: after an MMSBM has been fitted, can one perform a finite-sample goodness-of-fit test for residual directed structure that is not explained by the fitted dyad-level block-pair counts?

Goodness-of-fit testing for network models is difficult because the analyst typically observes one network rather than repeated independent networks.  The problem is especially delicate for sparse directed networks, where asymptotic approximations can be unreliable and the fitted model contains both latent variables and nuisance parameters.  Simulation-based diagnostics for network models often compare observed summaries with summaries generated from a fitted model, as in goodness-of-fit checks for exponential-family random graph models \parencite{hunter2008goodness}.  For block models, existing approaches include asymptotic goodness-of-fit tests for stochastic block models \parencite{lei2016goodness}, Monte Carlo tests for degree-corrected and related stochastic block models \parencite{karwa2024monte}, and spectral tests for stochastic block models \parencite{wuhu2024}.  Recent work on goodness-of-fit and testing for other discrete or network-supported data settings includes tests for discrete response models with covariates \parencite{meintanis2025goodness} and nonparametric tests for point processes on linear networks \parencite{gonzalezperez2025nonparametric}.  These methods are valuable, but they do not directly give a finite-sample exact conditional test for a directed MMSBM with dyad-level mixed-membership roles.

The present approach follows the exact conditional testing tradition.  Fisher's exact test and its extensions condition on sufficient margins to remove nuisance parameters \parencite{fisher1935design,mehta1983network,agresti1992survey}.  Generalized Monte Carlo significance tests give finite-sample valid Monte Carlo versions of such conditional tests \parencite{besag1989generalized}.  In algebraic statistics, the same idea is formulated in terms of fibers: one conditions on sufficient statistics and samples from all data configurations with the same sufficient-statistic value \parencite{diaconis1998algebraic}.  This viewpoint has also been used for exact conditional tests in network models \parencite{fienberg2011algebraic,gross2017goodness,ogawa2013graver}.

The key idea here is to condition on a sufficient statistic for the block probability matrix after fixing a latent block-pair assignment.  Once the block-pair edge totals are fixed, the nuisance block probabilities disappear and the conditional distribution is uniform over a finite fiber of binary networks.  This produces a finite-sample reference distribution for discrepancy statistics that are sensitive to structure not explained by the fitted block-pair totals.

The main contributions are as follows.  First, for a fixed latent block-pair assignment, we derive the exact conditional law of the network given the block-pair totals and show that it is uniform on a finite fiber.  Second, we construct a panel of fiber-sensitive discrepancies targeting sender and receiver heterogeneity, nodewise heterogeneity, reciprocity, and directed transitive triads.  These statistics are chosen because residual norms that depend only on block-pair totals are constant on the fiber and are therefore uninformative for conditional testing.  Third, we prove finite-sample validity of Monte Carlo rank \(p\)-values and of a Bonferroni omnibus test.  Fourth, we show how the test can be used when the block-pair assignment is latent by drawing a block-pair assignment from the posterior distribution and then applying the fixed-assignment conditional test.  Finally, we exploit the product structure of the fiber to obtain an independent uniform sampler, while also recording a simple swap chain that is useful for checking and for possible extensions.

The exactness statements are conditional and model-based.  They hold for a fixed block-pair assignment with exact uniform fiber sampling, and for the Bayesian MMSBM when the block-pair assignment used in the test is drawn exactly from its posterior distribution.  The numerical experiments therefore separate two regimes.  A fixed-label experiment uses the true simulated block-pair assignment and the direct fiber sampler, isolating the exact conditional calibration claim.  A second experiment assesses the practical procedure when the latent assignment is approximated by posterior simulation.  This separation identifies which part of the computation is responsible for any finite-sample size distortion.

The remainder of the paper is organized as follows.  Section \ref{sec:model} defines the directed MMSBM and the block-pair representation.  Section \ref{sec:fibre-framework} derives the conditional fiber law.  Section \ref{sec:discrepancies} introduces the discrepancy statistics.  Section \ref{sec:pvalues-section} proves validity of the conditional tests, including the version based on a posterior draw of the block-pair assignment.  Section \ref{sec:sampling} describes fiber sampling and the computational implementation.  Sections \ref{sec:simulation} and \ref{sec:sampson} present simulations and an empirical illustration, and Section \ref{sec:discussion} concludes.
\section{Model and notation}
\label{sec:model}

\subsection{Directed binary network setup}
\label{subsec:network-setup}

Let \(D\in\mathbb{N}\) denote the number of nodes.  We observe a directed binary network through its adjacency matrix
\[
Y=(Y_{ij})_{1\le i,j\le D}\in\{0,1\}^{D\times D},
\]
where \(Y_{ij}=1\) indicates a directed edge from node \(i\) to node \(j\), and \(Y_{ij}=0\) indicates its absence.  Self-loops are excluded throughout, so \(Y_{ii}=0\) for all \(i=1,\dots,D\).

Let
\[
\mathcal{D}:=\{(i,j):1\le i,j\le D,\ i\neq j\},\qquad |\mathcal{D}|=D(D-1)=:r,
\]
be the set of ordered dyads.  When convenient, we identify the adjacency matrix with a vector \(y\in\{0,1\}^r\) obtained by fixing an arbitrary ordering of the dyads in \(\mathcal{D}\).  We continue to write \(y_{ij}\) for the coordinate corresponding to the dyad \((i,j)\in\mathcal{D}\).  Thus \(Y\) and \(y\) contain the same information; the vector notation is used only to describe fibers and sampling distributions.

\subsection{Mixed membership stochastic block model}
\label{subsec:mmsbm}

Fix the number of latent groups \(K\ge 1\).  In a mixed membership model, each node may participate in more than one latent group.  This is represented by a membership vector
\[
\pi_i=(\pi_{i1},\dots,\pi_{iK})\in\Delta^{K-1},\qquad i=1,\dots,D,
\]
where
\[
\Delta^{K-1}:=\left\{x=(x_1,\dots,x_K)\in[0,1]^K:\sum_{k=1}^K x_k=1\right\}
\]
is the \((K-1)\)-dimensional probability simplex.  The component \(\pi_{ik}\) is the probability that node \(i\) acts in latent group \(k\) on a given interaction.

For each ordered dyad \((i,j)\in\mathcal{D}\), the model draws two latent roles:
\[
z^{(L)}_{ij}\in\{1,\dots,K\},\qquad z^{(R)}_{ij}\in\{1,\dots,K\}.
\]
Here \(z^{(L)}_{ij}\) is the role used by node \(i\) as the sender in the dyad \((i,j)\), while \(z^{(R)}_{ij}\) is the role used by node \(j\) as the receiver.  Conditional on the membership vectors, these dyad-level roles are independent across dyads and satisfy
\[
z^{(L)}_{ij}\mid \pi_i \sim \operatorname{Categorical}(\pi_i),
\qquad
z^{(R)}_{ij}\mid \pi_j \sim \operatorname{Categorical}(\pi_j).
\]
Equivalently,
\[
\mathbb{P}\{z^{(L)}_{ij}=k,\ z^{(R)}_{ij}=l\mid \pi_i,\pi_j\}
=\pi_{ik}\pi_{jl}.
\]

Let
\[
B=(B_{kl})_{k,l=1}^K\in(0,1)^{K\times K}
\]
be the block probability matrix.  The entry \(B_{kl}\) is the probability of a directed edge when the sender acts in group \(k\) and the receiver acts in group \(l\).  Conditional on the roles and on \(B\), the dyads are independent and
\[
Y_{ij}\mid z^{(L)}_{ij}=k,\ z^{(R)}_{ij}=l,\ B
\sim \operatorname{Bernoulli}(B_{kl}),\qquad (i,j)\in\mathcal{D}.
\]

For the version based on posterior-sampled block-pair assignments, we use the usual Bayesian version of the MMSBM.  The membership vectors and block probabilities are assigned independent priors,
\[
\pi_i \stackrel{\mathrm{iid}}{\sim} \operatorname{Dirichlet}(\alpha),\qquad i=1,\dots,D,
\]
for fixed \(\alpha\in(0,\infty)^K\), and
\[
B_{kl}\stackrel{\mathrm{ind}}{\sim}\operatorname{Beta}(\alpha_{kl},\beta_{kl}),
\qquad 1\le k,l\le K,
\]
for fixed hyperparameters \((\alpha_{kl},\beta_{kl})\).  These priors are used to obtain posterior draws of the latent roles.  The fixed-label conditional tests developed below depend on the Bernoulli likelihood after the dyad-level roles have been fixed, and not on the numerical values of the prior hyperparameters.

\subsection{Block-pair representation and block-pair totals}
\label{subsec:blockpair}

The conditional construction in this paper depends on the ordered pair of latent roles attached to each dyad.  It is useful to encode this ordered pair as a single block-pair label
\[
g_{ij}:=(z^{(L)}_{ij}-1)K+z^{(R)}_{ij}\in\{1,\dots,K^2\},
\qquad (i,j)\in\mathcal{D}.
\]
Thus each value of \(g_{ij}\) corresponds to exactly one ordered role pair \((k,l)\), with \(k\) the sender role and \(l\) the receiver role.  The full block-pair assignment is
\[
g=(g_{ij})_{(i,j)\in\mathcal{D}}.
\]

For each block pair \(h\in\{1,\dots,K^2\}\), define the set of dyads assigned to that block pair by
\[
\mathcal{E}_h:=\{(i,j)\in\mathcal{D}: g_{ij}=h\},
\qquad
n_h:=|\mathcal{E}_h|.
\]
The sets \(\mathcal{E}_1,\dots,\mathcal{E}_{K^2}\) form a partition of the ordered dyad set \(\mathcal{D}\).  The corresponding block-pair edge totals are
\[
m_h:=\sum_{(i,j)\in\mathcal{E}_h}Y_{ij},
\qquad h=1,\dots,K^2,
\]
and we write
\[
m=(m_1,\dots,m_{K^2}).
\]
These totals count how many observed edges occur inside each ordered block-pair class.  They are the sufficient statistics for the block probability matrix once the block-pair assignment \(g\) is fixed, and they will define the conditional fiber in Section \ref{sec:fibre-framework}.

In applications the dyad-level roles, and hence the block-pair assignment \(g\), are unobserved.  The conditional theory in Sections \ref{sec:fibre-framework}-\ref{sec:pvalues-section} is first developed for a fixed realization of \(g\).  Later, in Section \ref{subsec:posterior-draws}, we use posterior draws of \(g\) from a fitted MMSBM to propagate latent-role uncertainty.
\section{Exact conditional fiber-based goodness-of-fit testing}
\label{sec:fibre-framework}

\subsection{The conditional fiber law}
\label{subsec:conditional-fiber-law}

Fix a realization of the block-pair assignment $g$.  The sets
\[
\mathcal{E}_h=\{(i,j)\in\mathcal{D}:g_{ij}=h\},\qquad h=1,\dots,K^2,
\]
partition the ordered dyad set $\mathcal{D}$.  Within a fixed block pair $h$, all dyads share the same Bernoulli edge probability.  Therefore, once $g$ is fixed, the only information about the block probability matrix contained in the edges assigned to $\mathcal{E}_h$ is the number of observed edges in that class.

Recall that
\[
m_h=\sum_{(i,j)\in\mathcal{E}_h}Y_{ij},\qquad h=1,\dots,K^2,
\]
and write $m=(m_1,\dots,m_{K^2})$.  The block-pair fiber determined by $(g,m)$ is the set of all directed binary networks with the same block-pair edge totals as the observed network:
\begin{equation}
\label{eq:fibre}
\mathcal{F}(m;g)
:=
\Big\{y\in\{0,1\}^r:\ \sum_{(i,j)\in\mathcal{E}_h} y_{ij}=m_h
\ \text{for all }h=1,\dots,K^2\Big\}.
\end{equation}
Thus two networks in $\mathcal{F}(m;g)$ may differ in the locations of their edges, but they agree on how many edges occur in every ordered block-pair class.  Conditioning on this fiber removes the nuisance block probabilities and leaves only the residual arrangement of edges within block pairs.

For a block-pair index $h\in\{1,\dots,K^2\}$, write $h=(k-1)K+l$ for the corresponding ordered role pair $(k,l)$ and set $p_h=B_{kl}$.

\begin{prop}[Block-pair totals are sufficient for $B$ given $g$]
\label{prop:sufficiency}
Fix $K$ and a realization of $g$.  Conditional on $(g,B)$, the likelihood of $Y$ depends on $Y$ only through the block-pair totals
\[
m=(m_h)_{h=1}^{K^2}.
\]
Equivalently, $m$ is sufficient for $B$ given $g$.
\end{prop}

\begin{proof}
Conditional on $(g,B)$, dyads are independent, and each dyad in $\mathcal{E}_h$ has Bernoulli parameter $p_h$.  Hence, for $y\in\{0,1\}^r$,
\[
\mathbb{P}(Y=y\mid g,B)
=
\prod_{h=1}^{K^2}\prod_{(i,j)\in\mathcal{E}_h}
p_h^{y_{ij}}(1-p_h)^{1-y_{ij}}.
\]
Collecting the terms within each block-pair class gives
\[
\mathbb{P}(Y=y\mid g,B)
=
\prod_{h=1}^{K^2}
p_h^{m_h(y)}(1-p_h)^{n_h-m_h(y)},
\]
where
\[
m_h(y):=\sum_{(i,j)\in\mathcal{E}_h}y_{ij}.
\]
The likelihood therefore depends on $y$ only through
\[
(m_1(y),\dots,m_{K^2}(y)).
\]
The Neyman-Fisher factorization theorem gives the stated sufficiency.
\end{proof}

\begin{corollary}[Uniform conditional law on the fiber]
\label{cor:uniform-fibre}
Fix a realization of $g$ and block-pair totals $m$.  For any $B\in(0,1)^{K\times K}$,
\[
\mathbb{P}(Y=y\mid g,m,B)
=
\left(\prod_{h=1}^{K^2}\binom{n_h}{m_h}^{-1}\right)
\mathbf{1}\{y\in\mathcal{F}(m;g)\}.
\]
The same conditional law is obtained after integrating out $B$ under independent Beta priors.
\end{corollary}

\begin{proof}
If $y\in\mathcal{F}(m;g)$, then $m_h(y)=m_h$ for every $h$.  By Proposition \ref{prop:sufficiency},
\[
\mathbb{P}(Y=y\mid g,B)
=
\prod_{h=1}^{K^2}p_h^{m_h}(1-p_h)^{n_h-m_h},
\]
which is constant over all $y\in\mathcal{F}(m;g)$.  Thus, conditional on $(g,m,B)$, every element of the fiber receives the same probability.

It remains only to count the elements of the fiber.  For each $h$, one must choose exactly $m_h$ edge positions among the $n_h$ dyads in $\mathcal{E}_h$.  These choices are independent across block-pair classes, so
\[
|\mathcal{F}(m;g)|
=
\prod_{h=1}^{K^2}\binom{n_h}{m_h}.
\]
This gives the stated uniform conditional distribution.

If $B$ is integrated out under independent Beta priors, the marginal likelihood is a product of Beta-binomial factors, each depending on $y$ only through $m_h(y)$.  Hence the same conditioning argument gives the same uniform law on $\mathcal{F}(m;g)$.
\end{proof}

The corollary is the basis of the proposed test.  It says that, after fixing the latent block-pair assignment and conditioning on the corresponding edge totals, the MMSBM makes no further distinction among networks in the same fiber.  Any systematic extremeness of the observed network relative to this uniform conditional distribution is therefore evidence of structure not explained by the fitted block-pair probabilities.

\subsection{Conditional goodness-of-fit tests}
\label{subsec:conditional-tests}

Let $\pi_{g,m}$ denote the uniform distribution on $\mathcal{F}(m;g)$.  For any statistic
\[
T:\mathcal{F}(m;g)\to\mathbb{R},
\]
the conditional null induced by the MMSBM is
\[
H_0(g,m):\quad Y\sim \pi_{g,m}.
\]
A conditional goodness-of-fit test compares the observed value $T(y^{(0)})$ with the distribution of $T(Y)$ when $Y$ is sampled uniformly from $\mathcal{F}(m;g)$.

This formulation has two useful features.  First, the reference distribution is free of the nuisance block probabilities $B$, because those probabilities have been removed by conditioning on the sufficient statistic $m$.  Second, the test is finite-sample and conditional on the observed block-pair totals; it does not require an asymptotic approximation for the distribution of the statistic.

The choice of $T$ is important.  A statistic that depends only on the block-pair totals is constant on $\mathcal{F}(m;g)$ and therefore cannot be used to test the conditional null.  Informative discrepancies must vary under rearrangements of edges within block-pair classes.  The next section introduces such statistics for sender and receiver heterogeneity, nodewise degree concentration, reciprocity, and directed transitive triads.

\section{Fiber-sensitive discrepancy statistics}
\label{sec:discrepancies}

Throughout this section, fix $(g,m)$ and define
\[
p_h:=\frac{m_h}{n_h},\qquad h=1,\dots,K^2,
\]
with the convention that degenerate block pairs with $n_h=0$ contribute no terms.
Let $\varepsilon>0$ be a small stabilizer.
In implementations one may also impose a minimum cell size $n_{\min}$ and omit terms with small denominators; this affects only the statistic, not the conditional null distribution.

\subsection{Sender and receiver heterogeneity within block pairs}

For block pair $h$ and sender $i$, define
\[
\mathcal{E}^{\mathrm{out}}_{i,h}:=\{(i,j)\in\mathcal{E}_h\},\qquad
n^{\mathrm{out}}_{i,h}:=|\mathcal{E}^{\mathrm{out}}_{i,h}|,
\qquad
d^{\mathrm{out}}_{i,h}:=\sum_{(i,j)\in\mathcal{E}^{\mathrm{out}}_{i,h}}Y_{ij}.
\]
The within-block-pair sender discrepancy is
\begin{equation}
\label{eq:Tout}
T_{\mathrm{out}}
:=
\sum_{h=1}^{K^2}\sum_{i=1}^D
\mathbf{1}\{n^{\mathrm{out}}_{i,h}\ge n_{\min}\}
\frac{\big(d^{\mathrm{out}}_{i,h}-n^{\mathrm{out}}_{i,h}p_h\big)^2}
{n^{\mathrm{out}}_{i,h}p_h(1-p_h)+\varepsilon}.
\end{equation}

Similarly, for block pair $h$ and receiver $j$, define
\[
\mathcal{E}^{\mathrm{in}}_{j,h}:=\{(i,j)\in\mathcal{E}_h\},\qquad
n^{\mathrm{in}}_{j,h}:=|\mathcal{E}^{\mathrm{in}}_{j,h}|,
\qquad
d^{\mathrm{in}}_{j,h}:=\sum_{(i,j)\in\mathcal{E}^{\mathrm{in}}_{j,h}}Y_{ij},
\]
and
\begin{equation}
\label{eq:Tin}
T_{\mathrm{in}}
:=
\sum_{h=1}^{K^2}\sum_{j=1}^D
\mathbf{1}\{n^{\mathrm{in}}_{j,h}\ge n_{\min}\}
\frac{\big(d^{\mathrm{in}}_{j,h}-n^{\mathrm{in}}_{j,h}p_h\big)^2}
{n^{\mathrm{in}}_{j,h}p_h(1-p_h)+\varepsilon}.
\end{equation}
These statistics detect sender- or receiver-side concentration that remains after the block-pair edge counts have been fixed.

\subsection{Nodewise sender and receiver heterogeneity}

Define the out-degree and in-degree
\[
d^{\mathrm{out}}_i:=\sum_{j\neq i}Y_{ij},
\qquad
d^{\mathrm{in}}_j:=\sum_{i\neq j}Y_{ij}.
\]
Under the conditional null, the block-pair mean assigned to dyad $(i,j)$ is $p_{ij}:=p_{g_{ij}}$.
Set
\[
\mu^{\mathrm{out}}_i:=\sum_{j\neq i}p_{ij},
\qquad
v^{\mathrm{out}}_i:=\sum_{j\neq i}p_{ij}(1-p_{ij}),
\]
and define
\begin{equation}
\label{eq:Toutnode}
T_{\mathrm{out,node}}
:=
\sum_{i=1}^D \frac{(d^{\mathrm{out}}_i-\mu^{\mathrm{out}}_i)^2}{v^{\mathrm{out}}_i+\varepsilon}.
\end{equation}
Analogously, set
\[
\mu^{\mathrm{in}}_j:=\sum_{i\neq j}p_{ij},
\qquad
v^{\mathrm{in}}_j:=\sum_{i\neq j}p_{ij}(1-p_{ij}),
\]
and define
\begin{equation}
\label{eq:Tinnode}
T_{\mathrm{in,node}}
:=
\sum_{j=1}^D \frac{(d^{\mathrm{in}}_j-\mu^{\mathrm{in}}_j)^2}{v^{\mathrm{in}}_j+\varepsilon}.
\end{equation}
These nodewise statistics aggregate deviations across block pairs and are useful when degree heterogeneity is distributed across several latent role combinations.

\subsection{Reciprocity and directed transitive triads}

The reciprocity statistic is
\begin{equation}
\label{eq:Trec}
T_{\mathrm{rec}}
:=\sum_{1\le i<j\le D}\mathbf{1}\{Y_{ij}=1,\ Y_{ji}=1\}.
\end{equation}
The directed transitive-triad statistic is
\begin{equation}
\label{eq:Ttri}
T_{\mathrm{tri}}
:=\sum_{\substack{i,j,k\in\{1,\dots,D\}\\ \text{distinct}}}Y_{ij}Y_{jk}Y_{ik}.
\end{equation}
For $T_{\mathrm{out}}$, $T_{\mathrm{in}}$, $T_{\mathrm{out,node}}$, and $T_{\mathrm{in,node}}$, we use two-sided tests because unusually small as well as unusually large heterogeneity can indicate lack of fit.
For $T_{\mathrm{rec}}$ and $T_{\mathrm{tri}}$, we use upper-tail tests.

A useful practical point is that many residual norms are constant on $\mathcal{F}(m;g)$ when the fitted mean is constant within each block pair.
Such statistics cannot distinguish the observed network from its conditional randomizations.
The discrepancies above were chosen because they change under within-block-pair rearrangements and therefore remain informative after conditioning.

\section{Conditional \texorpdfstring{$p$}{p}-values and finite-sample validity}
\label{sec:pvalues-section}

The previous sections identify the conditional null distribution: for fixed \((g,m)\), the network is uniformly distributed on the fiber \(\mathcal{F}(m;g)\).  This section explains how that conditional law is turned into valid goodness-of-fit tests.  In principle, one could enumerate the entire fiber and compute the exact tail probability of a discrepancy statistic.  In realistic networks the fiber is usually too large for enumeration, so we use independent samples from the uniform fiber law.  The resulting \(p\)-values are based on the rank of the observed statistic among the observed network and the sampled fiber networks.  This rank-based construction gives finite-sample conservativeness for any number of Monte Carlo draws.

\subsection{Monte Carlo \texorpdfstring{$p$}{p}-values from the conditional fiber}
\label{subsec:pvalues}

Fix \((g,m)\) and let \(\pi_{g,m}\) denote the uniform distribution on \(\mathcal{F}(m;g)\).  Let \(y^{(0)}\) be the observed network, written in vector form, and let
\[
y^{(1)},\dots,y^{(M)}\stackrel{\mathrm{iid}}{\sim}\pi_{g,m}
\]
be independent fiber samples.  For a statistic \(T:\mathcal{F}(m;g)\to\mathbb{R}\), the relevant tail depends on the alternative being tested.  If unusually large values of \(T\) indicate lack of fit, we use the upper-tail value
\begin{equation}
\label{eq:p-upper}
p^{+}(T)
:=
\frac{1+\sum_{b=1}^{M}\mathbf{1}\{T(y^{(b)})\ge T(y^{(0)})\}}{M+1}.
\end{equation}
If unusually small values are relevant, we use the lower-tail value
\[
p^{-}(T):=
\frac{1+\sum_{b=1}^{M}\mathbf{1}\{T(y^{(b)})\le T(y^{(0)})\}}{M+1}.
\]
For discrepancies where both tails are meaningful, we use
\begin{equation}
\label{eq:p-two}
p^{\pm}(T):=
\min\{1,\ 2\min\{p^{+}(T),p^{-}(T)\}\}.
\end{equation}
The numerator and denominator in \eqref{eq:p-upper} reflect the fact that the observed network is treated as one member of the same exchangeable collection as the \(M\) simulated networks.  This prevents zero Monte Carlo \(p\)-values and gives a conservative test even when \(M\) is finite.  For example, if no simulated network is as extreme as the observed one in the upper tail, the reported value is \(1/(M+1)\) rather than zero.

\begin{theorem}[Finite-sample validity of Monte Carlo rank \texorpdfstring{$p$}{p}-values]
\label{thm:mc-valid}
Fix \((g,m)\) and suppose that
\[
Y^{(0)},Y^{(1)},\dots,Y^{(M)}
\]
are independent draws from \(\pi_{g,m}\).  For any real-valued statistic \(T\), define \(p^{+}(T)\) by \eqref{eq:p-upper} and \(p^{\pm}(T)\) by \eqref{eq:p-two}.  Then, for every \(\alpha\in[0,1]\),
\[
\mathbb{P}\{p^{+}(T)\le \alpha\}\le \alpha,
\qquad
\mathbb{P}\{p^{\pm}(T)\le \alpha\}\le \alpha.
\]
Thus both \(p^{+}(T)\) and \(p^{\pm}(T)\) are conservative exact conditional \(p\)-values under \(H_0(g,m)\).
\end{theorem}

\begin{proof}
Under \(H_0(g,m)\), the variables
\[
T(Y^{(0)}),T(Y^{(1)}),\dots,T(Y^{(M)})
\]
are exchangeable.  For the upper-tail statistic define
\[
R:=1+\sum_{b=1}^M\mathbf{1}\{T(Y^{(b)})\ge T(Y^{(0)})\}.
\]
Then \(p^{+}(T)=R/(M+1)\).  The quantity \(R\) is the upper rank of the observed statistic among the \(M+1\) exchangeable statistic values, with ties counted in the conservative direction.  Conditional on the multiset of statistic values, exchangeability implies that the observed index is equally likely to be any of the \(M+1\) positions.  With conservative tie handling, the upper rank is therefore stochastically no smaller than a discrete uniform random variable on \(\{1,\dots,M+1\}\).  Hence
\[
\mathbb{P}\{p^{+}(T)\le\alpha\}\le\alpha.
\]
The same argument applied to \(-T\) gives the corresponding lower-tail statement for \(p^{-}(T)\).  Finally,
\[
\{p^{\pm}(T)\le\alpha\}
\subseteq
\{p^{+}(T)\le\alpha/2\}\cup\{p^{-}(T)\le\alpha/2\},
\]
and the union bound gives
\[
\mathbb{P}\{p^{\pm}(T)\le\alpha\}
\le
\mathbb{P}\{p^{+}(T)\le\alpha/2\}
+
\mathbb{P}\{p^{-}(T)\le\alpha/2\}
\le \alpha.
\]
\end{proof}

\subsection{Combining several discrepancies}
\label{subsec:omnibus}

The proposed method is diagnostic: different statistics are intended to detect different residual features of the network.  A reciprocity statistic, for example, is sensitive to mutual dyads, while a sender-heterogeneity statistic is sensitive to concentration of outgoing edges.  In applications it is therefore natural to report componentwise \(p\)-values.  At the same time, it is useful to have a single omnibus test that rejects when any member of the discrepancy panel is unusually extreme.

Let \(\mathcal{T}\) be a finite collection of discrepancy statistics.  For each \(T\in\mathcal{T}\), compute a valid conditional \(p\)-value \(p(T)\) under \(\pi_{g,m}\) using the appropriate tail convention.  We use the Bonferroni omnibus value
\begin{equation}
\label{eq:omnibus-bonf}
p_{\mathrm{omni}}
:=
\min\Big\{1,\ |\mathcal{T}|\min_{T\in\mathcal{T}}p(T)\Big\}.
\end{equation}
This construction is deliberately conservative.  Its advantage is that it does not require independence, or even weak dependence, among the component statistics.

\begin{prop}[Omnibus validity under the conditional null]
\label{prop:omnibus}
If every \(p(T)\) is super-uniform under \(\pi_{g,m}\), then \(p_{\mathrm{omni}}\) in \eqref{eq:omnibus-bonf} is also super-uniform under \(\pi_{g,m}\).  Consequently, the rule \(p_{\mathrm{omni}}\le\alpha\) controls the conditional familywise error rate at level \(\alpha\) under \(H_0(g,m)\).
\end{prop}

\begin{proof}
For any \(\alpha\in[0,1]\),
\[
\{p_{\mathrm{omni}}\le \alpha\}
\subseteq
\bigcup_{T\in\mathcal{T}}
\left\{p(T)\le\frac{\alpha}{|\mathcal{T}|}\right\}.
\]
The union bound and the super-uniformity of the component \(p\)-values imply
\[
\mathbb{P}\{p_{\mathrm{omni}}\le\alpha\}
\le
\sum_{T\in\mathcal{T}}
\mathbb{P}\left\{p(T)\le\frac{\alpha}{|\mathcal{T}|}\right\}
\le \alpha.
\]
\end{proof}

\subsection{Using posterior draws of the block-pair assignment}
\label{subsec:posterior-draws}

The conditional test above treats the block-pair assignment \(g\) as fixed.  In applications, however, the dyad-level roles are not observed, and hence \(g\) is unknown.  We handle this by drawing one plausible block-pair assignment from the posterior distribution.  After observing \(Y\), draw
\[
\widetilde G\sim p(g\mid Y),
\]
compute the corresponding block-pair totals \(m(Y,\widetilde G)\), and apply the fixed-\(g\) conditional test on the fiber determined by
\[
\big(\widetilde G,m(Y,\widetilde G)\big).
\]

The justification is simple.  Under the Bayesian MMSBM described in Section \ref{sec:model}, drawing \(\widetilde G\) from the posterior distribution after observing \(Y\) reproduces the same joint distribution as the unobserved model-generated pair \((Y,G)\).  Therefore, if the conditional test is valid when \(g\) is fixed at the model-generated value, it remains valid when \(g\) is replaced by an exact posterior draw.

\begin{theorem}[Validity after a posterior draw of the block-pair assignment]
\label{thm:posterior-sampled-valid}
Assume that the Bayesian MMSBM described in Section \ref{sec:model} is the data-generating model.  Let \(G\) be the latent block-pair assignment and let \(M=m(Y,G)\).  Suppose that \(p_{\mathrm{cond}}(Y,G)\) is a valid conditional \(p\)-value for the fixed-\(G\) fiber test, in the sense that
\[
\mathbb{P}\{p_{\mathrm{cond}}(Y,G)\le\alpha\mid G,M\}\le \alpha,
\qquad 0\le \alpha\le 1.
\]
Let \(\widetilde G\) be an exact draw from the posterior distribution \(p(g\mid Y)\).  Then
\[
\mathbb{P}\{p_{\mathrm{cond}}(Y,\widetilde G)\le\alpha\}\le \alpha,
\qquad 0\le \alpha\le 1.
\]
Thus a test that first draws \(\widetilde G\) from \(p(g\mid Y)\) and then applies the fixed-\(g\) conditional test has level at most \(\alpha\).
\end{theorem}

\begin{proof}
For any network \(y\) and block-pair assignment \(g\),
\[
\mathbb{P}(Y=y,\widetilde G=g)
=
\mathbb{P}(Y=y)\mathbb{P}(\widetilde G=g\mid Y=y).
\]
Since \(\widetilde G\) is drawn from the posterior distribution of \(G\),
\[
\mathbb{P}(\widetilde G=g\mid Y=y)
=
\mathbb{P}(G=g\mid Y=y).
\]
Therefore
\[
\mathbb{P}(Y=y,\widetilde G=g)
=
\mathbb{P}(Y=y)\mathbb{P}(G=g\mid Y=y)
=
\mathbb{P}(Y=y,G=g).
\]
Hence \((Y,\widetilde G)\) and \((Y,G)\) have the same joint distribution.

It remains to use the fixed-\(G\) conditional validity.  By Corollary \ref{cor:uniform-fibre}, given \((G,M)\), the network \(Y\) is uniformly distributed on \(\mathcal{F}(M;G)\).  Therefore,
\[
\mathbb{P}\{p_{\mathrm{cond}}(Y,G)\le\alpha\mid G,M\}\le \alpha.
\]
Taking expectations over \((G,M)\) gives
\[
\mathbb{P}\{p_{\mathrm{cond}}(Y,G)\le\alpha\}\le \alpha.
\]
Since \((Y,\widetilde G)\) and \((Y,G)\) have the same joint distribution, the same bound holds with \(\widetilde G\) in place of \(G\):
\[
\mathbb{P}\{p_{\mathrm{cond}}(Y,\widetilde G)\le\alpha\}\le \alpha.
\]
\end{proof}
\section{Sampling from the fiber}
\label{sec:sampling}

The conditional tests developed above require samples from the uniform distribution on \(\mathcal{F}(m;g)\).  Throughout this section, the pair \((g,m)\) is treated as fixed.  When the block-pair assignment is unknown, one first obtains a posterior draw of \(g\) from the fitted MMSBM and then computes the corresponding totals \(m(Y,g)\).  After this step, the sampling problem is purely conditional: sample uniformly from the fiber determined by the fixed pair \((g,m)\).

The main point is that this fiber has a simple product structure.  The constraints defining \(\mathcal{F}(m;g)\) are imposed separately within each block-pair class \(\mathcal{E}_h\).  Hence, to draw a network from the fiber, one only has to choose which \(m_h\) of the \(n_h\) dyads in each block-pair class receive edges.  These choices can be made independently across block pairs.  This gives an exact independent sampler and avoids the burn-in and autocorrelation issues that arise in Markov-chain sampling.

\subsection{Direct uniform sampling from the fiber}
\label{subsec:direct-sampler}

For each block pair \(h\in\{1,\dots,K^2\}\), choose a subset
\[
S_h\subseteq \mathcal{E}_h
\]
of size \(m_h\) uniformly among all \(\binom{n_h}{m_h}\) such subsets, independently over \(h\).  Then set
\[
y_{ij}=1 \quad \text{if } (i,j)\in S_h \text{ for its block pair } h,
\]
and set \(y_{ij}=0\) otherwise.

\begin{prop}[Direct uniform sampler]
\label{prop:direct-sampler}
The construction above generates an independent draw from the uniform distribution on \(\mathcal{F}(m;g)\).
\end{prop}

\begin{proof}
Every element \(y\in\mathcal{F}(m;g)\) corresponds uniquely to the collection of subsets
\[
S_h(y)=\{(i,j)\in\mathcal{E}_h:y_{ij}=1\},\qquad h=1,\dots,K^2,
\]
with \(|S_h(y)|=m_h\) for every \(h\).  Conversely, any such collection of subsets defines a unique element of the fiber.  The sampler chooses each \(S_h\) with probability \(\binom{n_h}{m_h}^{-1}\), independently across block-pair classes.  Hence each \(y\in\mathcal{F}(m;g)\) is assigned probability
\[
\prod_{h=1}^{K^2}\binom{n_h}{m_h}^{-1}
=
|\mathcal{F}(m;g)|^{-1},
\]
and states outside the fiber have probability zero.  Therefore the resulting draw is uniform on \(\mathcal{F}(m;g)\).
\end{proof}

This direct sampler is the default implementation used in the numerical work.  It is exact for the conditional fiber distribution and produces independent Monte Carlo samples.  The only Monte Carlo error in the conditional step is therefore ordinary independent-sampling error.

\subsection{Implementation of the conditional test}
\label{subsec:algorithm}

Algorithm \ref{alg:full} summarizes the implementation.  For each posterior draw of the block-pair assignment, the algorithm computes the corresponding block-pair totals, samples networks from the associated fiber, and evaluates the chosen discrepancy statistics on the observed and sampled networks.  A single posterior draw gives the test covered by Theorem \ref{thm:posterior-sampled-valid}.  Using several posterior draws is useful in applications because it shows how sensitive the diagnostic conclusions are to uncertainty in the latent block-pair assignment.

\begin{algorithm}[H]
\caption{Conditional goodness-of-fit testing using posterior draws of block-pair assignments}
\label{alg:full}
\begin{algorithmic}[1]
\Require Directed network \(Y\in\{0,1\}^{D\times D}\) with \(Y_{ii}=0\); number of groups \(K\); prior hyperparameters; number of posterior draws \(U\); conditional Monte Carlo size \(M\); discrepancy set \(\mathcal{T}\); stabilizer \(\varepsilon\); minimum cell size \(n_{\min}\).
\Ensure Componentwise conditional \(p\)-values \(p_u(T)\) for \(u=1,\dots,U\) and \(T\in\mathcal{T}\); optionally, omnibus values \(p_{u,\mathrm{omni}}\).
\State Fit the MMSBM with \(K\) groups to \(Y\) and obtain posterior draws \(g^{(1)},\dots,g^{(U)}\).
\For{\(u=1,\dots,U\)}
  \State Compute the block-pair sets \(\mathcal{E}^{(u)}_h=\{(i,j)\in\mathcal{D}:g^{(u)}_{ij}=h\}\), \(h=1,\dots,K^2\).
  \State Compute the corresponding block-pair totals \(m^{(u)}_h=\sum_{(i,j)\in\mathcal{E}^{(u)}_h}Y_{ij}\), \(h=1,\dots,K^2\).
  \State Let \(y^{(0)}\) be the fixed vectorization of the observed network \(Y\). Compute \(T(y^{(0)})\) for all \(T\in\mathcal{T}\).
  \For{\(b=1,\dots,M\)}
     \State Draw \(y^{(b)}\) uniformly from \(\mathcal{F}(m^{(u)};g^{(u)})\) using Proposition \ref{prop:direct-sampler}.
     \State Compute \(T(y^{(b)})\) for all \(T\in\mathcal{T}\).
  \EndFor
  \State Compute \(p_u(T)\) using \eqref{eq:p-upper} or \eqref{eq:p-two}, according to the tail convention for \(T\).
  \State Optionally compute \(p_{u,\mathrm{omni}}\) using \eqref{eq:omnibus-bonf}.
\EndFor
\State \Return \(\{p_u(T):u=1,\dots,U,\ T\in\mathcal{T}\}\) and, if requested, \(\{p_{u,\mathrm{omni}}:u=1,\dots,U\}\).
\end{algorithmic}
\end{algorithm}

For the \(u\)th posterior draw, the size of the conditional fiber is
\[
|\mathcal{F}(m^{(u)};g^{(u)})|
=
\prod_{h=1}^{K^2}\binom{n^{(u)}_h}{m^{(u)}_h}.
\]
The quantity
\[
\sum_{h=1}^{K^2}\log\binom{n^{(u)}_h}{m^{(u)}_h}
\]
is useful as a diagnostic.  If this log fiber size is very small, then the conditional distribution has little variation, and some discrepancy statistics may have limited ability to distinguish the observed network from its fiber randomizations.  When the fiber is large, the direct sampler provides many independent rearrangements of the observed block-pair edge counts.

With the direct sampler, Monte Carlo uncertainty can be summarized using ordinary binomial standard errors for tail counts.  Repeating the calculation over several posterior draws of \(g\) gives a separate assessment of sensitivity to the latent block-pair assignment.

\subsection{A swap chain for checking and extensions}
\label{subsec:swap-chain}

The direct sampler above is preferable for the basic conditioning scheme used in this paper.  Nevertheless, it is useful to record a Markov-chain sampler on the same fiber.  The chain gives a simple way to verify implementations, supports incremental computation, and may become useful if additional constraints are imposed in future extensions, because such constraints can destroy the simple product structure used by the direct sampler.

Fix \((g,m)\) and write \(\mathcal{F}=\mathcal{F}(m;g)\).  Let
\[
\mathcal{H}:=\{h\in\{1,\dots,K^2\}:0<m_h<n_h\}
\]
be the set of nondegenerate block pairs.  A swap move chooses a block pair \(h\in\mathcal{H}\), selects one current \(1\)-position and one current \(0\)-position within \(\mathcal{E}_h\), and exchanges their values.  Let \((q_h)_{h\in\mathcal{H}}\) be positive probabilities with \(\sum_{h\in\mathcal{H}}q_h=1\).  At each step, sample \(h\sim q\), choose the current \(1\)-position and \(0\)-position uniformly within \(\mathcal{E}_h\), and swap them.  A lazy version, which holds with probability \(1/2\), may be used to remove periodicity.

\begin{prop}[Reversibility and stationarity]
\label{prop:swap-reversible}
The swap chain preserves \(\mathcal{F}(m;g)\) and is reversible with respect to the uniform distribution on \(\mathcal{F}(m;g)\).  Therefore the uniform distribution on the fiber is stationary.
\end{prop}

\begin{proof}
A swap within block pair \(h\) leaves the total \(m_h\) unchanged and does not alter any other block-pair total, so the chain remains in the fiber.  Let \(y,y'\in\mathcal{F}(m;g)\) be two distinct states connected by one swap in block pair \(h\).  Conditional on selecting \(h\), the probability of proposing that swap is
\[
\frac{1}{m_h(n_h-m_h)},
\]
because there are \(m_h\) ones and \(n_h-m_h\) zeros in \(\mathcal{E}_h\).  The reverse transition has the same probability.  Multiplying by \(q_h\) gives \(P(y,y')=P(y',y)\).  Since the target distribution is uniform, detailed balance holds.
\end{proof}

\begin{lemma}[Swap connectivity]
\label{lem:swap-connects}
For any \(y,y'\in\mathcal{F}(m;g)\), a finite sequence of within-block-pair swaps transforms \(y\) into \(y'\) while remaining in \(\mathcal{F}(m;g)\).
\end{lemma}

\begin{proof}
The constraints are block-pairwise, so it is enough to treat one block pair at a time.  Fix \(h\) and enumerate
\[
\mathcal{E}_h=\{e_1,\dots,e_{n_h}\}.
\]
Let \(u,u'\in\{0,1\}^{n_h}\) be the restrictions of \(y\) and \(y'\) to these coordinates.  Both have exactly \(m_h\) ones.  Define
\[
A:=\{a:u_a=1,\ u'_a=0\},
\qquad
B:=\{b:u_b=0,\ u'_b=1\}.
\]
Because \(u\) and \(u'\) have the same number of ones, \(|A|=|B|\).  If \(A\) is empty, the two restrictions already agree.  Otherwise, choose \(a\in A\) and \(b\in B\), and swap the entries at \(a\) and \(b\).  This reduces the Hamming distance between \(u\) and \(u'\) by two while preserving the block-pair total.  Repeating the argument completes the transformation within block pair \(h\), and concatenating the transformations over all \(h\) proves the result.
\end{proof}

\begin{remark}[Relation to conditional sampling in algebraic statistics]
Exact conditional tests for discrete models are often formulated in terms of fibers: one conditions on sufficient statistics and samples over all data configurations with the same sufficient-statistic value.  For general log-linear models, connecting such fibers may require a Markov basis, namely a finite set of moves that is guaranteed to connect every fiber \parencite{diaconis1998algebraic,fienberg2011algebraic,gross2017goodness}.  In the present MMSBM conditioning problem, the fiber is much simpler.  Once \(g\) and \(m\) are fixed, the fiber factorizes into independent fixed-sum binary strings over the block-pair classes \(\mathcal{E}_h\).  Lemma \ref{lem:swap-connects} shows that ordinary \(1/0\) swaps within each block pair already connect the whole fiber.  Thus no general Markov-basis computation is needed for the conditional sampler used here.
\end{remark}
\section{Simulation study}
\label{sec:simulation}

The simulation study has three aims.  The first is to check the finite-sample calibration guaranteed by the conditional theory when the latent block-pair assignment is fixed and known.  The second is to assess the implemented procedure when the block-pair assignment is unknown and replaced by a posterior draw from a fitted MMSBM.  The third is to study power and diagnostic localization under directed alternatives that introduce residual structure beyond the fitted block-pair totals.

The results support the main claims of the paper.  With the true block-pair assignment fixed, the proposed conditional test is close to nominal under the null.  With posterior-drawn block-pair assignments, the implemented procedure remains mildly conservative in the null experiments.  Under alternatives, the method is particularly effective for directed reciprocity and transitive closure, while the sender-hub experiment reveals a limitation of the present discrepancy panel.

\subsection{Design}
\label{subsec:simulation-design}

We simulate directed binary networks from an MMSBM with \(K_{\mathrm{true}}=3\) latent groups.  The node-level mixed-membership vectors are sampled independently from a Dirichlet distribution with common concentration parameter \(0.5\).  We consider two network sizes, \(D\in\{20,30\}\), and two sparsity regimes.  The block probabilities are generated from $\operatorname{Beta}(1,\beta),\beta\in\{4,9\}$
corresponding to mean baseline edge probabilities \(1/(1+\beta)\in\{0.20,0.10\}\).  Thus the design includes one moderately dense and one sparse directed setting.

We consider four data-generating scenarios:
\begin{enumerate}[leftmargin=2.1em,itemsep=0.25em]
\item \textit{Null}: the network is generated directly from the directed MMSBM.
\item \textit{Reciprocity alternative}: after an MMSBM draw, isolated one-way dyads are reciprocated with additional probability \(0.25\).
\item \textit{Triadic-closure alternative}: after an MMSBM draw, repeated sweeps add edges of the form \(i\to k\) when \(i\to j\) and \(j\to k\) are already present.
\item \textit{Sender-hub alternative}: each node receives an additional sender-side logit shift, creating outgoing heterogeneity not explained by the fitted block structure.
\end{enumerate}

The fixed-label calibration experiment uses \(500\) independent null networks for each design cell.  In this experiment we condition on the true simulated block-pair assignment and sample directly from the corresponding fiber using Proposition \ref{prop:direct-sampler}.  This isolates the exact conditional part of the method, since there is no posterior uncertainty in \(g\) and no Markov-chain error in the fiber sampling step.

The experiment with posterior-drawn block-pair assignments uses \(300\) independent networks for each design cell.  For each dataset we fit the MMSBM with \(K=3\) groups, obtain posterior draws of the block-pair assignment by collapsed Gibbs sampling, and compute the omnibus conditional \(p\)-value using one posterior-drawn assignment.  Fiber samples are again drawn by the direct sampler.  Thus the remaining numerical approximations come from posterior simulation of the block-pair assignment and from finite conditional Monte Carlo sampling.

\subsection{Fixed-label exact calibration}
\label{subsec:fixed-label-calibration}

\begin{figure}[H]
\centering
\includegraphics[width=.88\textwidth]{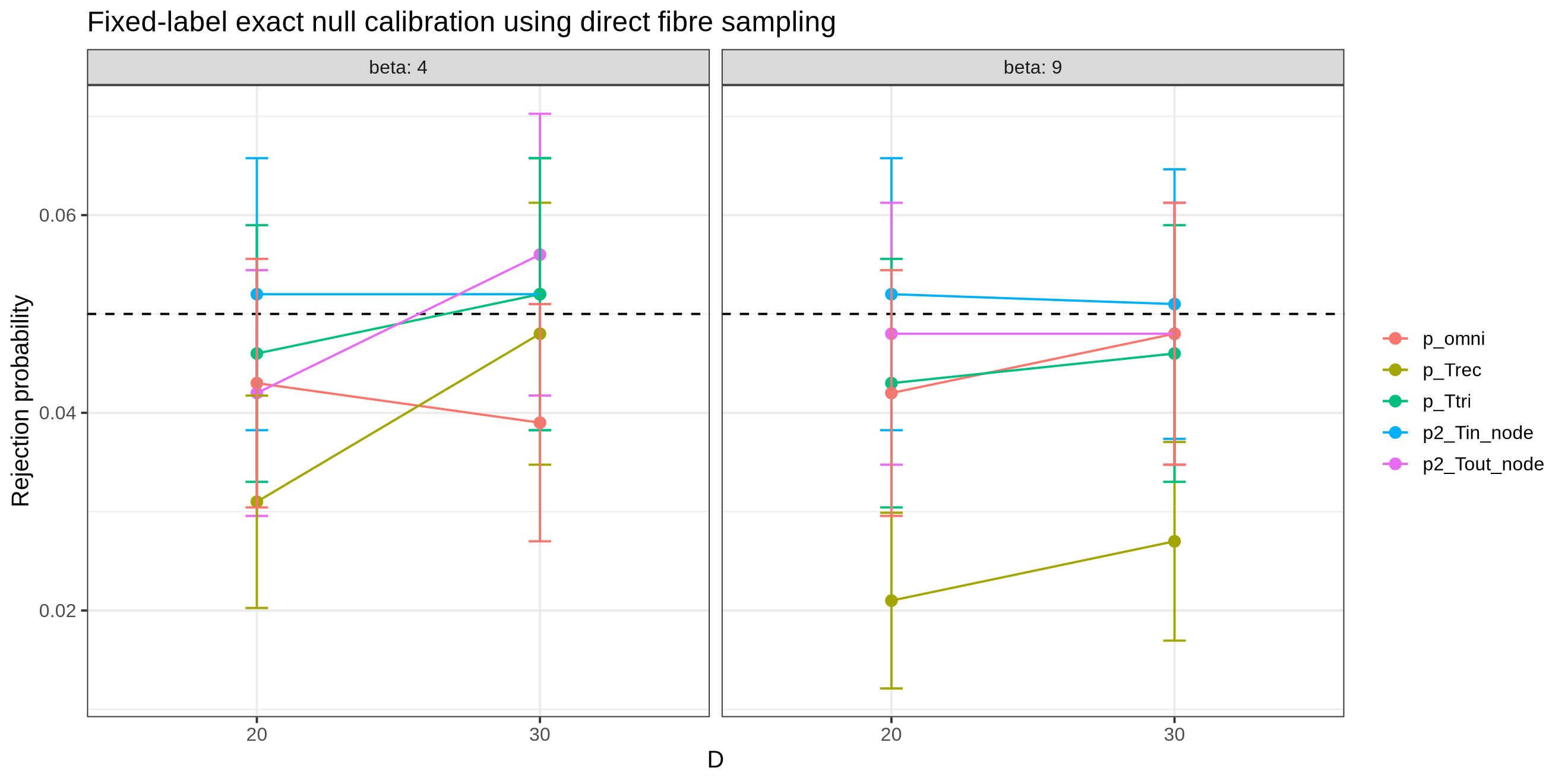}
\caption{Fixed-label exact null calibration using direct fiber sampling. The latent block-pair assignment is fixed at its true simulated value. Points show rejection rates at nominal level \(0.05\), with Monte Carlo confidence intervals. The dashed line marks the nominal level.}
\label{fig:fixed-label-calibration}
\end{figure}

Figure \ref{fig:fixed-label-calibration} reports rejection rates under the fixed-label null.  The proposed omnibus test is close to the nominal level \(0.05\) in all four design cells.  The rejection rates are \(0.042\), \(0.036\), \(0.042\), and \(0.048\) for \((D,\beta)=(20,4),(20,9),(30,4),(30,9)\), respectively.  The componentwise tests show the same behavior: averaged over the four design cells, their rejection rates range from \(0.0345\) for the reciprocity statistic to \(0.0510\) for the nodewise sender statistic.

These results confirm the finite-sample calibration predicted by the conditional theory.  When the conditioning variables are fixed and the fiber is sampled directly, the proposed test behaves as an exact conditional test should.  The slight conservativeness of the omnibus statistic is expected because the Bonferroni combination is conservative and because \(p_{\mathrm{omni}}\) is truncated at one.

\subsection{Benchmark methods}
\label{subsec:benchmarks}

We compare the proposed method with two SBM-oriented goodness-of-fit procedures:
\begin{enumerate}[leftmargin=2.1em,itemsep=0.25em]
\item the Monte Carlo beta-SBM test of \textcite{karwa2024monte};
\item the spectral SBM test based on the linear spectral statistic of \textcite{wuhu2024}.
\end{enumerate}
These benchmarks are designed for undirected hard-assignment block models, whereas the proposed method is designed for directed mixed-membership networks.  To give the benchmarks a reasonable input, each directed network is OR-symmetrized and hard labels are obtained by spectral clustering into \(K=3\) groups.  The comparison should therefore be interpreted as a practical benchmark comparison rather than a like-for-like comparison of methods under a common null.  For the beta-SBM benchmark, a small number of numerical failures occur; all reported rates for that method are computed using the nonmissing outputs.

\subsection{Calibration with posterior-drawn block-pair assignments}
\label{subsec:calibration}

\begin{figure}[H]
\centering
\includegraphics[width=.88\textwidth]{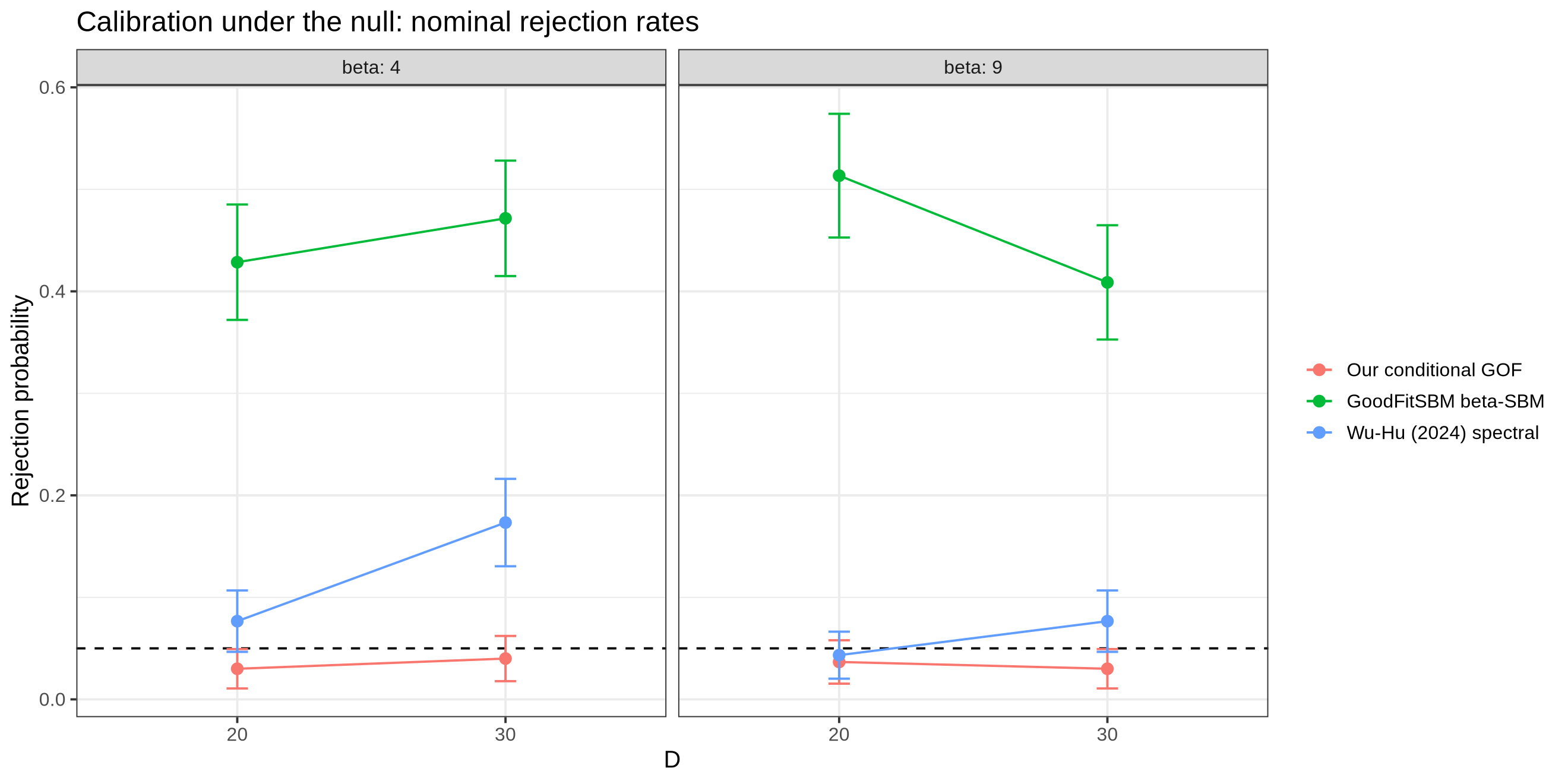}
\caption{Nominal null rejection rates for the proposed method, the beta-SBM benchmark, and the Wu-Hu spectral benchmark. The dashed line marks the nominal level \(0.05\).}
\label{fig:sim-calibration}
\end{figure}

Figure \ref{fig:sim-calibration} reports null rejection rates for the implemented procedure with posterior-drawn block-pair assignments and for the two benchmarks.  The proposed omnibus test remains close to nominal and is mildly conservative.  The null rejection rates are \(0.030\), \(0.0367\), \(0.040\), and \(0.030\) for \((D,\beta)=(20,4),(20,9),(30,4),(30,9)\), respectively.

This shows that the practical implementation is well calibrated in the simulated MMSBM settings considered here.  The result is important because it goes beyond the ideal fixed-label experiment: the block-pair assignment is no longer fixed at its true simulated value, but is instead obtained from posterior simulation.

The benchmark results reflect the mismatch between their intended setting and the directed mixed-membership data-generating mechanism.  The beta-SBM procedure has substantially larger null rejection rates in these experiments.  The Wu-Hu spectral benchmark is closer to nominal, except in the moderately dense \(D=30\) case.  This does not indicate a failure of those methods in their intended hard-assignment SBM settings; rather, it shows that their calibration is not targeted to the directed MMSBM setting studied here.

\subsection{Power under directed alternatives}
\label{subsec:alternative-comparison}

Because the empirical sizes differ across methods, we compare power using size-adjusted rejection rates.  For each method and each design cell \((D,\beta)\), we estimate the empirical \(5\%\) null cutoff from the corresponding null simulations and evaluate rejection under each alternative relative to that cutoff.  This separates sensitivity to alternatives from size distortion under the null.

\begin{figure}[H]
\centering
\includegraphics[width=.95\textwidth]{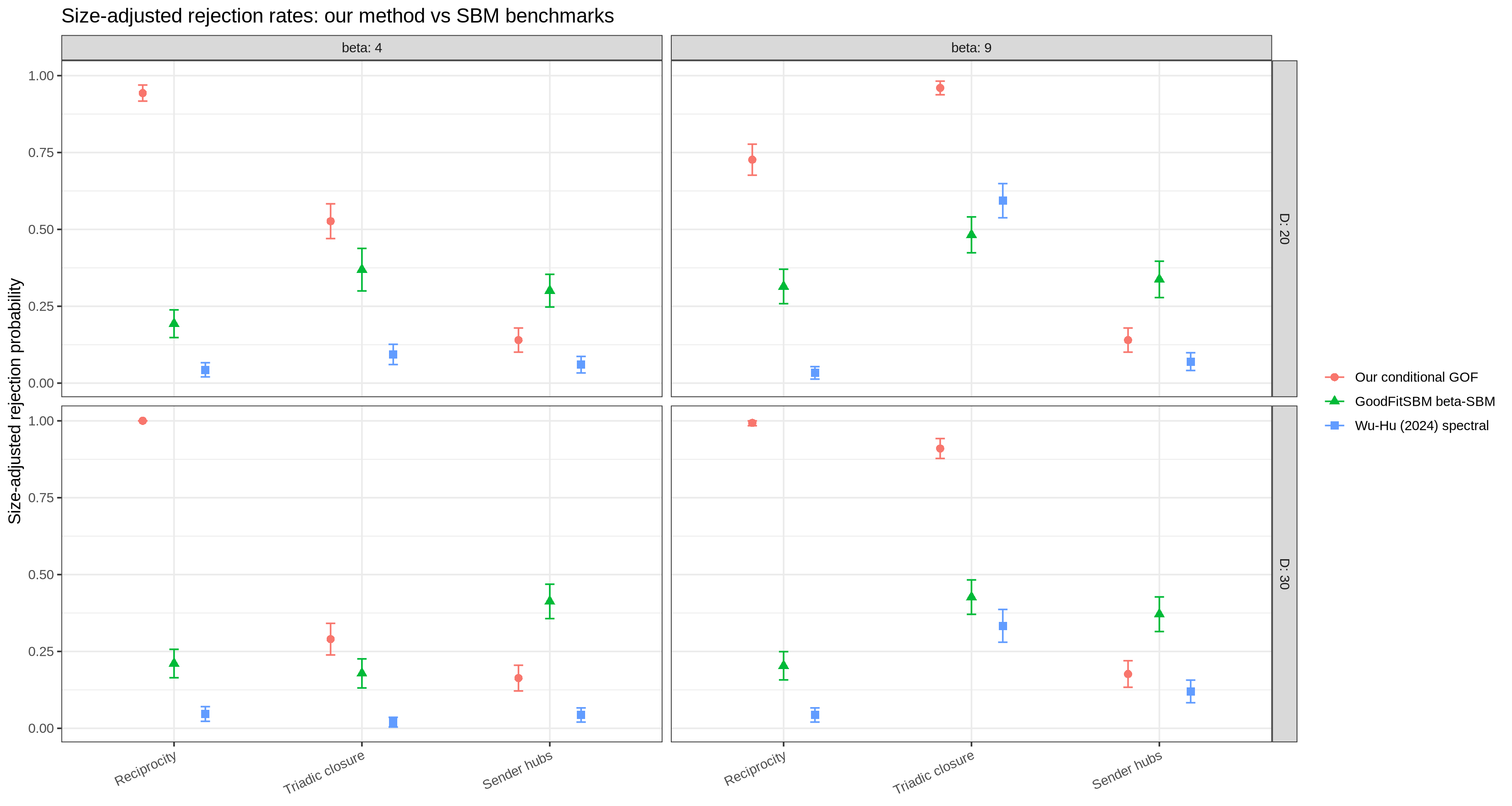}
\caption{Size-adjusted rejection rates under the reciprocity, sender-hub, and triadic-closure alternatives. The proposed method is applied to the original directed network, whereas the SBM benchmarks are applied to a symmetrized hard-label reduction.}
\label{fig:sim-bench-adjusted}
\end{figure}

Figure \ref{fig:sim-bench-adjusted} gives the size-adjusted results.  The proposed method is strongest under the reciprocity alternative.  Averaged over the four \((D,\beta)\) design cells, the proposed omnibus rejection rate is \(0.916\), compared with \(0.230\) for the beta-SBM benchmark and \(0.042\) for the Wu-Hu spectral benchmark.  This is the clearest advantage of the directed conditional approach.  The reciprocal signal is a directed feature, and it is weakened or lost when the network is reduced to an undirected hard-label representation.

The proposed method is also effective under triadic closure.  The average size-adjusted omnibus rejection rate is \(0.672\), compared with \(0.364\) for beta-SBM and \(0.260\) for Wu-Hu.  The advantage is especially pronounced in the sparse regime \(\beta=9\), where the proposed omnibus test rejects with rates \(0.960\) for \(D=20\) and \(0.910\) for \(D=30\).  Thus conditioning on block-pair totals does not remove the higher-order directed structure created by transitive closure.

The sender-hub alternative is less favorable to the proposed omnibus statistic.  The average rejection rate is \(0.155\), compared with \(0.355\) for beta-SBM and \(0.073\) for Wu-Hu.  This is a useful limitation rather than a contradiction of the method.  Sender hubs primarily create degree heterogeneity, and this may also be visible after symmetrization.  The present discrepancy panel is most sensitive to directed residual structure such as mutuality and transitivity.  If sender effects are the primary alternative of interest, stronger degree-oriented discrepancies should be added to the panel.

\subsection{Summary of simulation evidence}
\label{subsec:simulation-summary}

The simulation study gives four conclusions.  First, the fixed-label experiment confirms the finite-sample calibration predicted by the conditional theory.  Second, the implementation with posterior-drawn block-pair assignments remains mildly conservative under the correctly specified MMSBM in the reported null experiments.  Third, the proposed method has strong power against directed reciprocity and good power against directed transitive closure, especially in sparse networks.  Fourth, the componentwise \(p\)-values are diagnostically useful: they identify reciprocity and transitive closure as the source of lack of fit when those are the true alternatives.

The main limitation revealed by the simulations is the sender-hub experiment.  The conditional framework itself remains valid, but the present discrepancy panel is not equally sensitive to all types of residual structure.  In applications where sender- or receiver-specific degree effects are a central concern, the panel should be supplemented with stronger degree-oriented statistics.
\section{Empirical illustration}
\label{sec:sampson}

We illustrate the method on the directed Sampson monastery network \parencite{sampson1968}. The network has
$D=18$ nodes and directed edge density $0.183$. We take $K=3$ as the main MMSBM specification and use
$K\in\{2,3,4\}$ as a robustness check.

For the main $K=3$ fit, we used 300 posterior draws of the block-pair assignment. For each repetition, one
posterior block-pair assignment was drawn from the fitted MMSBM, the corresponding block-pair totals were
computed, and the conditional fiber distribution was sampled directly using Proposition 6.1
with $M=1500$ independent conditional Monte Carlo draws. Thus each repetition corresponds to one
posterior-draw conditional test of the kind justified in Theorem 5.3; the table below
summarizes the stability of these conditional p-values over repeated posterior draws. Because the fiber draws
are independent, no fiber-chain burn-in or thinning is required. The remaining numerical uncertainty comes
from posterior simulation of the latent block-pair assignment and from finite conditional Monte Carlo sampling.

\begin{table}[h]
\centering
\caption{Sampson network, main $K=3$ fit: summary of posterior-drawn conditional p-values over
300 repetitions. The last column reports the proportion of repeated posterior draws of the block-pair assignment
for which the corresponding conditional p-value is at most $0.05$.}
\label{tab:sampson-k3}
\begin{tabular}{lccc}
\toprule
Statistic & Mean p-value & Median p-value & Proportion $p\le 0.05$ \\
\midrule
$p^{\pm}(T_{\mathrm{out}})$      & 0.143 & 0.027 & 0.597 \\
$p^{\pm}(T_{\mathrm{in}})$       & 0.258 & 0.164 & 0.180 \\
$p^{\pm}(T_{\mathrm{out,node}})$ & 0.005 & 0.001 & 0.987 \\
$p^{\pm}(T_{\mathrm{in,node}})$  & 0.146 & 0.090 & 0.323 \\
$p^{+}(T_{\mathrm{rec}})$        & 0.040 & 0.030 & 0.717 \\
$p^{+}(T_{\mathrm{tri}})$        & 0.298 & 0.260 & 0.030 \\
$p_{\mathrm{omni}}$              & 0.013 & 0.004 & 0.950 \\
\bottomrule
\end{tabular}
\end{table}

The main $K=3$ fit shows strong evidence of lack of fit. The omnibus p-value is small for most posterior
draws of the block-pair assignment: its mean and median are $0.013$ and $0.004$, respectively, and
$95.0\%$ of the repeated posterior-draw tests have $p_{\mathrm{omni}}\le 0.05$.

The componentwise results localize the lack of fit. The strongest signal is nodewise sender heterogeneity:
$p^{\pm}(T_{\mathrm{out,node}})$ has mean p-value $0.005$ and falls below $0.05$ in $98.7\%$ of the repetitions.
Reciprocity is also a stable signal, with mean p-value $0.040$ and rejection proportion $71.7\%$. The
within-block-pair sender statistic provides a secondary but more variable signal; its median p-value is
$0.027$ and it falls below $0.05$ in $59.7\%$ of repetitions, even though its mean is larger because the
posterior-draw distribution is skewed.

Receiver-side discrepancies are less extreme. The within-block-pair receiver and nodewise receiver statistics
have mean p-values $0.258$ and $0.146$, respectively, with rejection proportions $18.0\%$ and $32.3\%$.
The directed transitive-triad statistic is not a systematic signal in this analysis: its mean and median p-values
are $0.298$ and $0.260$, and only $3.0\%$ of repetitions fall below $0.05$. After conditioning on fitted
block-pair totals, the residual structure in the Sampson network is therefore associated mainly with sender-side
heterogeneity and mutuality, rather than with receiver-side heterogeneity or directed transitive closure.

As a robustness check, we repeated the analysis for $K\in\{2,3,4\}$ in a separate $K$-grid run with 200
posterior-draw repetitions for each value of $K$. The omnibus conclusion remains stable, although larger
values of $K$ absorb some residual structure. For $K=2,3,4$, the mean omnibus p-values are $0.007$,
$0.022$, and $0.046$, the corresponding medians are $0.008$, $0.008$, and $0.016$, and the proportions
of repetitions with $p_{\mathrm{omni}}\le 0.05$ are $1.000$, $0.905$, and $0.700$. These $K=3$ robustness
numbers differ slightly from those in Table \ref{tab:sampson-k3} because the robustness run used 200
posterior-draw repetitions, whereas the main $K=3$ table used 300 repetitions.

The componentwise robustness results point to the same qualitative conclusion. Across $K=2,3,4$, the
nodewise sender statistic rejects in proportions $1.000$, $0.985$, and $0.935$, while the reciprocity statistic
rejects in proportions $1.000$, $0.720$, and $0.685$. The within-block-pair sender statistic becomes less
extreme as $K$ increases, with rejection proportions $0.850$, $0.510$, and $0.285$, suggesting that additional
latent groups absorb part of the local sender concentration. Nevertheless, sender-side heterogeneity and
reciprocity remain the dominant residual features across the fitted group sizes. The empirical lack of fit is
therefore not an artifact of a single choice of $K$.

\section{Discussion}
\label{sec:discussion}

We developed an exact conditional goodness-of-fit framework for the directed MMSBM.  For a fixed block-pair assignment, conditioning on the block-pair edge totals removes the nuisance block probability matrix and gives a uniform distribution on a finite fiber.  This conditional law supports finite-sample tests for discrepancies that vary within the fiber, including sender and receiver heterogeneity, reciprocity, and directed transitive closure.

The basic computation is simple because the fiber factorizes across block pairs and can be sampled directly.  The swap-chain formulation is mainly useful for checking, interpretation, and possible extensions with additional constraints.

The simulations support the main theoretical claims.  With the true block-pair assignment fixed, the test is close to nominal under the null.  With posterior-sampled block-pair assignments, the implemented method remains well calibrated in the reported null experiments.  Under alternatives, the method is especially effective for directed reciprocity and remains useful for directed transitive closure.  The Sampson analysis similarly finds residual sender-side and reciprocal structure not explained by the fitted MMSBM.

The main limitation is the choice of discrepancy panel.  The sender-hub simulations suggest that stronger degree-oriented statistics may be needed when node-level heterogeneity is the primary concern.  Another practical limitation is that posterior sampling of the latent block-pair assignment is approximate in applications.  Future work should develop faster posterior samplers, sharper diagnostics for latent-assignment uncertainty, and model extensions incorporating reciprocal or sender-effect structure while preserving tractable conditional inference.
\section*{Statement on AI usage}
The author used ChatGPT to generate the R code for the simulation study, help draft portions of the manuscript, and assist in verifying the mathematical results.

\printbibliography

\end{document}